\documentclass[11pt,twoside]{article}
\usepackage{asp2004}
\usepackage{epsf}
\usepackage{psfig}
\begin{document}

\title{Kinematics of the Ionized Halo of NGC 5775}
\author{George H. Heald, Richard J. Rand}
\affil{University of New Mexico, 800 Yale Blvd NE, Albuquerque NM 87131}
\author{Robert A. Benjamin}
\affil{University of Wisconsin, 800 W. Main St, Whitewater WI 53190}
\author{Joseph A. Collins}
\affil{University of Colorado, Campus Box 389, Boulder CO 80309}
\author{Joss Bland-Hawthorn}
\affil{Anglo-Australian Observatory, P.~O. Box 296, Epping NSW, Australia}

\begin{abstract}
Studies of the kinematics of gaseous halos in some spiral galaxies have shown that the rotational velocity of the gas decreases with height above the midplane ($z$). This vertical velocity gradient has been observed in both the neutral and ionized components of halos. We present imaging Fabry--Perot H$\alpha$ observations of the edge-on galaxy NGC 5775. These observations allow us to study the variation in the rotation curve (and, to some extent, the radial profile) of the ionized gas component as a function of $z$. We have used an iterative technique to infer the major axis rotation curve from the ionized gas observations as well as CO data. A preliminary analysis of position-velocity (PV) diagrams parallel to the major axis implies the presence of a vertical velocity gradient. The magnitude of this gradient is tentatively found to be $\sim 2$ km s$^{-1}$ arcsec$^{-1}$, consistent with the results of \citet{rand00} and \citet{tullmann00}.
\end{abstract}

\section{Introduction}

Deep observations of the gaseous halos in some spiral galaxies (e.g., NGC 891, \citeauthor{swaters97} \citeyear{swaters97}; UGC 7321, \citeauthor{matthews03} \citeyear{matthews03}; NGC 2403, \citeauthor{fraternali02} \citeyear{fraternali02}) have revealed a decrease in rotational velocity with height above the midplane ($z$). This decrease has been observed in both the neutral and diffuse ionized gas (DIG) halo components. The fountain model of \citet{bregman80} may explain this behavior because as gas is lifted into the halo, it feels a weaker gravitational potential, migrates radially outward, and thus its rotation speed drops in order to conserve angular momentum. Thus far, the global effects on the kinematics of an individual galaxy have been simulated by treating the fountain as a purely ballistic process \citep{collins02}, and as a purely hydrodynamic process (\citeauthor{benjamin00} \citeyear{benjamin00}; Ciotti, this volume). Further detail and comparisons with available data are provided by Rand \citetext{this volume}.

While \ion{H}{I} observations provide data that can be used to investigate the vertical velocity gradient in a spatially complete manner, complementary programs in the optical typically make use of slit spectroscopy. Therefore, two-dimensional velocity data for the DIG component are usually not available. A prime candidate for slit spectroscopy has been NGC 5775, because it is relatively nearby, nearly edge-on \citep[$D=24.8$ Mpc, $i=86$\deg;][]{irwin94}, and shows a remarkably thick and bright DIG layer with vertical filaments \citep{collins00}. \citet{rand00} found that the velocities measured along two slits parallel to the minor axis were consistent with a decrease in rotational velocity with $z$. \citet{tullmann00} found the same trend with a third slit, also parallel to the minor axis. In this paper, we present Fabry--Perot observations of the ionized gas in NGC 5775. These observations enable us to study the rotation speed as a function of $z$ in a manner similar to studies of the vertical velocity gradient in the neutral component of galaxy halos. We briefly describe the observations and data reductions in \S 2, describe an analysis of the velocity field in \S 3, and summarize in \S 4.

\section{Observations and Data Reduction}

Details of the observations and data reduction techniques will be described in detail in a forthcoming paper. For the sake of brevity, only the most important details are included here. Data were obtained during the nights of 11--13 April 2001 at the Anglo-Australian Observatory (AAT), using the TAURUS-II Fabry--Perot interferometer (order 379) in conjunction with the MIT/LL 2k$\times$4k CCD and an order blocking filter (6601/15). The spectral resolution is FWHM $\simeq$ 0.5\AA\ = 22.9 km s$^{-1}$, and the free spectral range (FSR) is 17.4\AA . Standard data reduction steps (bias and flat field corrections and cosmic ray removal) were performed using IRAF\footnote{IRAF is distributed by the National Optical Astronomy Observatories, which are operated by the Association of Universities for Research in Astronomy, Inc., under cooperative agreement with the National Science Foundation}. Sky-line removal, intensity and wavelength calibrations, construction of a data cube, and continuum subtraction were performed using scripts written in MATLAB. The final cube was smoothed to a spatial resolution of 3.67 arcseconds. The noise in the channel maps was measured to be $4.93 \times 10^{-18}$ erg cm$^{-2}$ s$^{-1}$ arcsec$^{-2}$ (assuming T = $10^4$K, this corresponds to an emission measure of 2.47 pc cm$^{-6}$).

\section{Analysis and Modeling}

In order to characterize the kinematics of the DIG component in the halo of NGC 5775, one can glean some basic trends from the velocity field (moment-1 map). However, as several authors \citep[e.g.][]{kregel04} have pointed out, this information is of limited use in the study of edge-on galaxies. For a given resolution element in the plane of the sky, the observed velocity profile contains contributions from every detectable location in the disk intercepted by the line of sight (LOS). In an edge-on galaxy, the result is that each velocity profile contains contributions from many different galactocentric radii ($R$). In general, the gas at these radii is characterized by different gas densities and different rotational velocities $v(R)$. It is clear, then, that the recovery of a rotation curve in an edge-on spiral is more complicated than simply calculating the mean velocity.

Several methods have been suggested for recovering the rotation curve of a galaxy. In particular, two useful methods in the study of edge-on spirals are the envelope tracing method (\citeauthor{sancisi79} \citeyear{sancisi79}; \citeauthor{sofue01} \citeyear{sofue01}), which calculates the rotation curve using the high-velocity edge of a position-velocity (PV) diagram, and the ``iteration method'' \citep{takamiya02}, which automates the procedure of generating a model galaxy (with specified rotation curve and density distribution) that best matches the kinematic properties of the observed galaxy. The benefit of using the iteration method is a more accurate recovery of the rotation curve at small radii, which, because of beam smearing and rapidly changing densities and velocities, is typically underestimated by the envelope tracing method.

We have written MATLAB scripts to implement the envelope tracing and iteration methods. Space constraints prohibit a description of the algorithms, which will be presented in detail in a forthcoming paper. We note, however, that optimizing the performance of the iteration method is critically dependent on the selection of an appropriate radial density profile. In addition, the signal-to-noise and velocity dispersion in the model galaxy must match those of the data. These requirements are imposed by the fact that the iteration method seeks to generate a model PV diagram which best matches a corresponding PV diagram derived from the data, using rotation curves generated from these diagrams by the envelope tracing method as a convergence criterion.

The envelope tracing and iteration methods work under the assumption that the disk is transparent. When working in the optical, this assumption is not necessarily valid. Indeed, optical images of NGC 5775 reveal a prominent dust lane running parallel to the major axis. While the location of the dust lane indicates that it may not interfere with calculating the rotation curve along the major axis, extinction is still a concern. With that in mind, we have calculated a major axis rotation curve using CO 2-1 data \citep{lee01} kindly provided by S.-W. Lee. We assume that the spatial and velocity distribution of CO and DIG is identical in the disk so that we can compare rotation curves derived separately from the two individual data sets.

Figure 1 shows the major axis rotation curve of NGC 5775, derived from the CO and H$\alpha$ data, using the iteration method. In both cases, the points were determined by clipping the velocity profiles at 3 times the rms noise in the PV diagram. Error bars were determined by clipping the velocity profiles at 2 and 4 times the rms noise and running the program again. The rotation curve determined by \citet{irwin94} for the neutral gas component is plotted for reference.

\begin{figure}[!ht]
\plotfiddle{fig1a.eps}{2.1cm}{0}{100}{100}{0}{0}
\plotfiddle{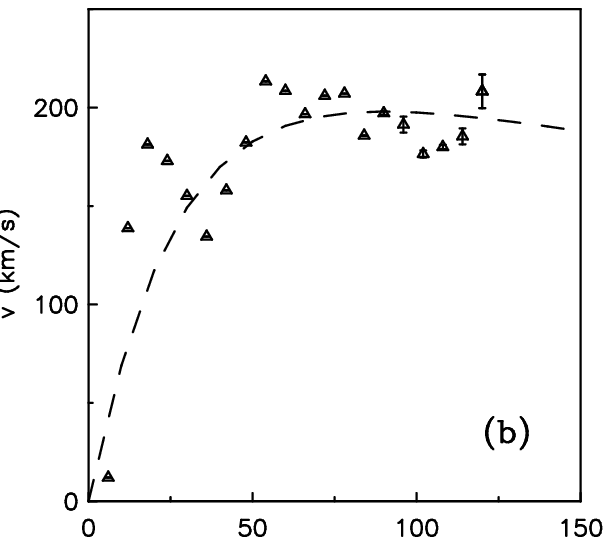}{2.1cm}{0}{100}{100}{0}{0}
\caption{Major axis rotation curve of NGC 5775, derived from (\emph{a}) CO data for the approaching side (\emph{crosses}) and receding side (\emph{circles}), and (\emph{b}) H$\alpha$ data for the approaching side (\emph{triangles}). The rotation curve from \citet{irwin94} is plotted for reference (\emph{dashed line}).}
\end{figure}

The CO data have a much larger beam (21\arcsec) than the H$\alpha$ data (3.67\arcsec). Thus, it should not be too surprising that the rotation curve derived from the CO data is much smoother. We only show the H$\alpha$ rotation curve obtained for the approaching side because the receding side rotation curve was found to be very low, perhaps due to extinction obscuring the gas on the line of nodes. Further analysis will be required to explain the kinematics of the gas on the receding side of the major axis.

\begin{figure}[!ht]
\plotfiddle{fig2a.eps}{3.4cm}{0}{100}{100}{0}{0}
\plotfiddle{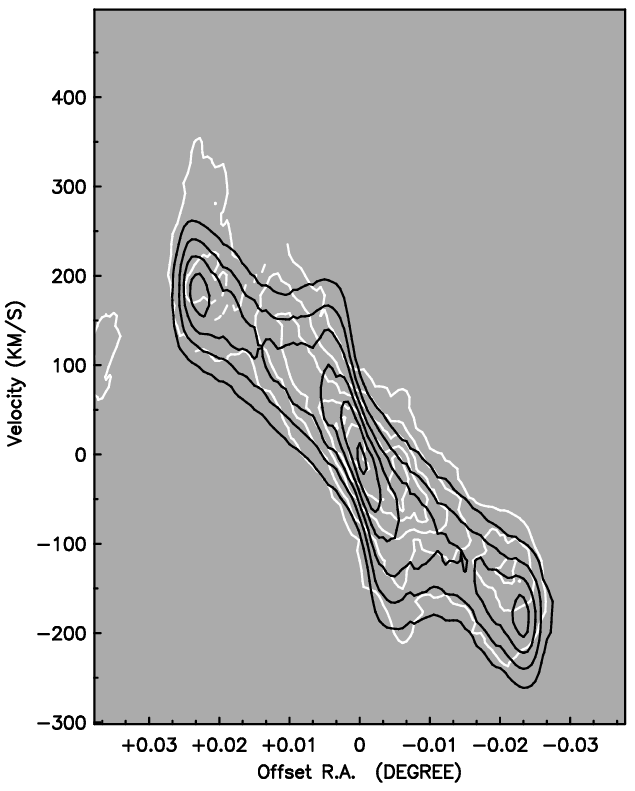}{3.4cm}{0}{100}{100}{0}{0}
\plotfiddle{fig2c.eps}{3.4cm}{0}{100}{100}{0}{0}
\plotfiddle{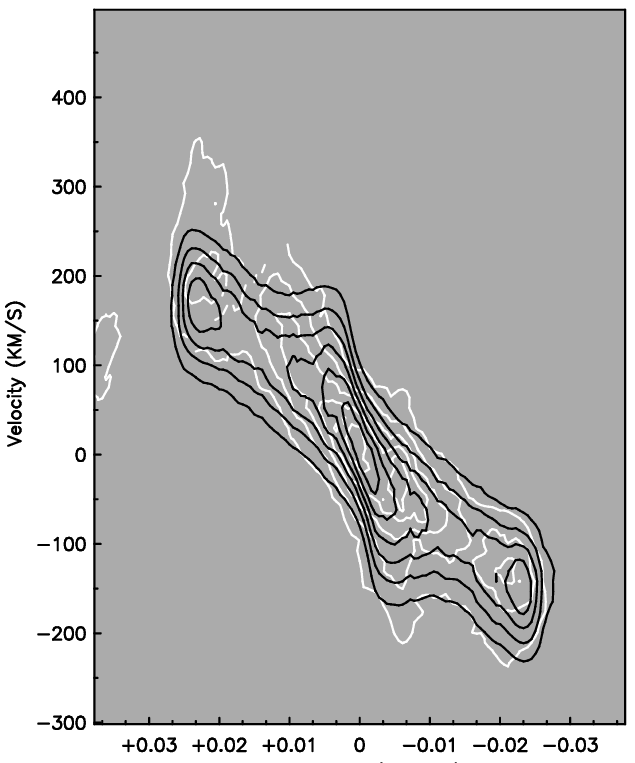}{3.4cm}{0}{100}{100}{0}{0}
\caption{Position-velocity diagrams from the H$\alpha$ data (white) and galaxy models (black). \emph{Upper left:} Major axis; contours run from 5 to 505$\times$rms, in increments of 50$\times$rms. \emph{Upper right:} $z = -20$ arcsec; contours run from 5 to 55$\times$rms, in increments of 5$\times$rms. \emph{Lower left:} $z = -20$ arcsec; same contour values; $v_{sys}$ has been offset by 20 km s$^{-1}$. \emph{Lower right:} $z = -20$ arcsec; same contour values; $v_{sys}$ has been offset by 10 km s$^{-1}$, and a vertical velocity gradient of 2 km s$^{-1}$ arcsec$^{-1}$ has been added.}
\end{figure}

To check for a dropoff in rotational velocity with $z$, we have modified the GIPSY task GALMOD to include a linear vertical velocity gradient:
\begin{equation}
v(R,z) = \left\{
\begin{array}{ll}
v(R,z=0) - \frac{dv}{dz}(|z|-z_0) & \mbox{for $z > z_0$}\nonumber\\
v(R,z=0) & \mbox{for $z \leq z_0$},
\end{array}
\right.
\end{equation}
where $z_0$ is the scale height of the vertical gas distribution, assumed to be exponential, and $dv/dz$ is a constant parameter in the model. We have generated galaxy models using this code, and visually compared model PV diagrams from the halo with corresponding diagrams obtained from the data. It is important to note that in this preliminary study, the models are constructed under the assumption that the shape of the radial density profile is constant with $z$. This assumption will not be true if halo gas is, for example, radially outflowing.

To begin the analysis, we have examined PV diagrams at $|z| = 20$ arcsec above the midplane, on both the NE and SW sides of the disk. This height was selected to eliminate concerns about extinction, and to avoid confusing the halo with the edge of the inclined disk. \citet{collins00} report that extinction effects should become negligible at $z \approx 600$ pc, or 5--6 arcsec. At an inclination of 86\deg, and assuming an optical radius of 120 arcsec, emission from gas at $z = 0$ (in the galaxy frame) should be negligible at projected $z$-heights greater than 10 arcsec.

In Figure 2, we present a sequence of PV diagrams. Note that the receding (Offset RA $>$ 0) side will not be well matched, as we have not yet derived a reliable rotation curve for that side. In the upper left panel, the major axis PV diagram shows the extent to which we were able to match the model (black) to the data (white) in the midplane. Next, in the upper right panel, the PV diagram from a slit parallel to the major axis, but offset by 20 arcsec to the SW, is displayed and compared to the same model. No vertical velocity gradient has been added. The inner part of the PV diagram from the data is shallower, and reaches lower velocities than are seen in the model. In the lower left panel, we have added 20 km s$^{-1}$ to the systemic velocity in the model. The inner region is much steeper in the model, but this may be explained by a change in the radial density profile (the radial profile and rotation curve are very difficult to decouple in the inner region). Therefore, we cannot yet rule out a shift in the systemic velocity of the halo. Finally, in the lower right panel, we display a model with a vertical velocity gradient of 2 km s$^{-1}$ arcsec$^{-1}$. The systemic velocity has been offset by 10 km s$^{-1}$ to achieve a better match. Again, a modification of the shape of the radial density profile may be required. It is not yet clear which combination of parameters produces the best model, but this final model appears most like the data. Results from the NE side of the major axis are similar in appearance.

\section{Summary and Future Work}

The clumpiness of the ionized gas distribution, and the possibility of a change in the radial density profile with height, makes it difficult to choose the best model. At this stage of our analysis, a vertical velocity gradient of 2 km s$^{-1}$ arcsec$^{-1}$ is consistent with the observed velocity data. However, an offset in the systemic velocity of the halo without a changing rotation curve appears to fit the data as well. Clearly, further analysis is in order.

As our analysis continues, we are examining PV diagrams along additional slices parallel to the major axis. The extra information from these slices may enable us to distinguish between a systemic velocity shift in the halo and the changing rotation curve predicted by the fountain model. Moreover, if fountain-type gas is flowing radially outward as it is lifted into the halo, the effects may be more easily recognized as a change in the radial distribution of gas with increasing $z$. This possibility will be explored. We may also be able to detect a lagging halo directly by deriving rotation curves for gas at $z > 0$.

\acknowledgments{This material is based on work parially supported by the National Science Foundation under Grant No. AST 99-86113. We thank Filippo Fraternali for the use of his GIPSY task. We are also grateful to Siow-Wang Lee for providing the CO 2-1 data cube, and Judith Irwin for providing an \ion{H}{I} cube.}


\begin{thebibliography}{}
\bibitem[Benjamin(2000)]{benjamin00}
Benjamin, R.~A. 2000, Rev. Mexicana Astron. Astrofis. Ser. Conf., 9, 256
\bibitem[Bregman(1980)]{bregman80}
Bregman, J.~N. 1980, ApJ, 236, 577
\bibitem[Collins et al.(2000)]{collins00}
Collins, J.~A., Rand, R.~J., Duric, N., \&\ Walterbos, R.~A.~M. 2000, ApJ, 536, 645
\bibitem[Collins, Benjamin, \&\ Rand(2002)]{collins02}
Collins, J.~A., Benjamin, R.~A., \&\ Rand, R.~J. 2002, ApJ, 578, 98
\bibitem[Fraternali et al.(2002)]{fraternali02}
Fraternali, F., van Moorsel, G., Sancisi, R., \&\ Oosterloo, T. 2002, AJ, 123, 3124
\bibitem[Irwin(1994)]{irwin94}
Irwin, J.~A. 1994, ApJ, 429, 618
\bibitem[Kregel \&\ van der Kruit(2004)]{kregel04}
Kregel, M. \&\ van der Kruit, P.~C. 2004, MNRAS, 352, 787
\bibitem[Lee et al.(2001)]{lee01}
Lee, S.-W., Irwin, J.~A., Dettmar, R.-J., Cunningham, C.~T., Golla, G., \&\ Wang, Q.~D. 2001, A\&A, 377, 759
\bibitem[Matthews \&\ Wood(2003)]{matthews03}
Matthews, L.~D. \&\ Wood, K. 2003, ApJ, 593, 721
\bibitem[Rand(2000)]{rand00}
Rand, R.~J. 2000, ApJ, 537, L13
\bibitem[Sancisi \&\ Allen(1979)]{sancisi79}
Sancisi, R. \&\ Allen, R.~J. 1979, A\&A, 74, 73
\bibitem[Sofue \&\ Rubin(2001)]{sofue01}
Sofue, Y. \&\ Rubin, V. 2001, ARAA, 39, 137
\bibitem[Swaters, Sancisi, \&\ van der Hulst(1997)]{swaters97}
Swaters, R.~A., Sancisi, R., \&\ van der Hulst, J.~M. 1997, ApJ, 491, 140
\bibitem[Takamiya \&\ Sofue(2002)]{takamiya02}
Takamiya, T. \&\ Sofue, Y. 2002, ApJ, 576, L15
\bibitem[T\"ullmann et al.(2000)]{tullmann00}
T\"ullmann, R., Dettmar, R.-J., Soida, M., Urbanik, M., \&\ Rossa, J. 2000, A\&A, 364, L36
\end{thebibliography}
\end{document}